\documentclass[pre,superscriptaddress,twocolumn,nofootinbib,floatfix]{revtex4}

\usepackage{amsmath,amsfonts,amssymb}
\usepackage{graphicx}
\usepackage{hyperref}
\usepackage{verbatim}
\usepackage{comment}

\usepackage[letterpaper, foot=10cm, left=0.75in, right=0.75in, top=0.75in, bottom=0.75in]{geometry}

\begin{document}
\thispagestyle{plain}

\title{Ring around the colloid}
\author{Marcello Cavallaro Jr.}
\thanks{These authors contributed equally to this work.}
\affiliation{Department of Chemical and Biomolecular Engineering, University of Pennsylvania, Philadelphia, Pennsylvania 19104, USA}
\author{Mohamed A. Gharbi}
\thanks{These authors contributed equally to this work.}
\affiliation{Department of Chemical and Biomolecular Engineering, University of Pennsylvania, Philadelphia, Pennsylvania 19104, USA}
\affiliation{Department of Physics and Astronomy, University of Pennsylvania, Philadelphia, Pennsylvania 19104, USA}
\affiliation{Department of Materials Science and Engineering, University of Pennsylvania, Philadelphia, PA 19104, USA}
\author{Daniel A. Beller}
\thanks{These authors contributed equally to this work.}
\affiliation{Department of Physics and Astronomy, University of Pennsylvania, Philadelphia, Pennsylvania 19104, USA}
\author{Simon \v{C}opar}
\affiliation{Department of Physics and Astronomy, University of Pennsylvania, Philadelphia, Pennsylvania 19104, USA}
\author{Zheng Shi}
\affiliation{Department of Chemistry, University of Pennsylvania, Philadelphia, Pennsylvania 19104, USA}
\author{Randall D. Kamien}
\affiliation{Department of Physics and Astronomy, University of Pennsylvania, Philadelphia, Pennsylvania 19104, USA}
\author{Shu Yang}
\affiliation{Department of Chemical and Biomolecular Engineering, University of Pennsylvania, Philadelphia, Pennsylvania 19104, USA}
\affiliation{Department of Materials Science and Engineering, University of Pennsylvania, Philadelphia, PA 19104, USA}
\author{Tobias Baumgart}
\affiliation{Department of Chemistry, University of Pennsylvania, Philadelphia, Pennsylvania 19104, USA}
\author{Kathleen J. Stebe}
\email{kstebe@seas.upenn.edu}
\affiliation{Department of Chemical and Biomolecular Engineering, University of Pennsylvania, Philadelphia, Pennsylvania 19104, USA}

\begin{abstract}
In this work, we show that Janus washers, genus-one colloids with hybrid anchoring conditions, form topologically required defects in nematic liquid crystals. Experiments under crossed polarizers reveal the defect structure to be a rigid disclination loop confined within the colloid, with an accompanying defect in the liquid crystal. When confined to a homeotropic cell, the resulting colloid-defect ring pair tilts relative to the far field director, in contrast to the behavior of toroidal colloids with purely homeotropic anchoring. We show that this tilting behavior can be reversibly suppressed by the introduction of a spherical colloid into the center of the toroid, creating a new kind of multi-shape colloidal assemblage.
\end{abstract}

\maketitle

\section*{}
\vspace{-1cm}

The term ``defect,'' even when dressed with the word ``topological,'' \cite{Mermin,Klemanbook} implies something to be avoided! Of course, this is not the case -- defects in crystals lead to stiffening, sometimes a desirable property \cite{hirth}; defects in cholesteric liquid crystals can be exploited to make bistable cholesteric displays \cite{kdi}; blue phases, rife with defects, are leading to the next generation of high speed video technologies \cite{bluephase}. With those and other lofty goals in mind, there has been a surge of interest in exquisitely controlling defects in liquid crystals to strive for new levels of self-organization  \cite{SmalyukhMCLC,Musevic2006,ravnik2011three,Nych3D}, hybrid materials \cite{RavnikJanus,KangOpals}, and novel phenomena \cite{Torons,Lasers}. The control of topological defects necessarily starts with the control of {\sl boundary conditions} or, equivalently, surface anchoring. Further, it has been shown that particle shape has pronounced effects on colloidal interactions and assembly \cite{GharbiLiquidCrystals, lapointe2009shape, dontabhaktuni2012shape}. From a slightly different perspective, disclinations in liquid crystals have been used as a template mechanism to direct the assembly of colloids into ordered patterns via elastic interactions \cite{Galerne2007, MusevicPRE2008}.

\begin{figure} [h]
\centering
\includegraphics[width=.48\textwidth]{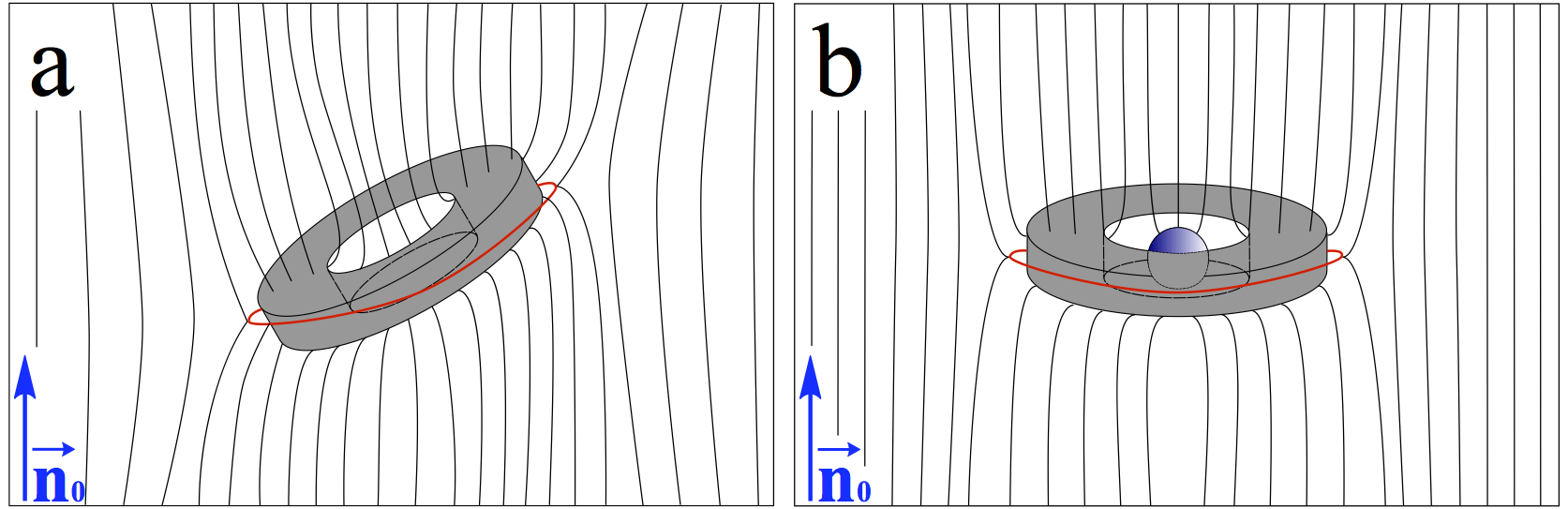}
\caption{(a) Schematic of an isolated hybrid washer in a uniform director field. With degenerate planar anchoring on the lower face and homeotropic anchoring on all other faces, the hybrid washer is topologically charged and is accompanied by a single disclination loop. (b) Schematic of a two-colloid assembly consisting of a homeotropic sphere inside a hybrid washer: a ring around the colloid. Due to the symmetry imposed by the sphere, the assembly orients perpendicular to the director field.}
\label{Fig1}
\end{figure}

The orientation, assembly, and topology of genus-zero particles, including spheres and rods, is now well understood. The defect structures around particles of higher genus, however, have only recently been explored \cite{topcolloids}. Here we examine the effects of both nontrivial anchoring and nontrivial colloid shape by preparing colloidal rings with hybrid boundary conditions to make ``Janus washers''. As shown in Fig.\ref{Fig1}a,
three sides of the washer are prepared with homeotropic (perpendicular) anchoring while the bottom has degenerate planar anchoring. We expect these washers to promote a winding of the director field by $\pi$ as we encircle the handle of the washer and so these ``hybrid washers'' will carry a nonvanishing topological charge in the 3D nematic \cite{RMP2}.  Compared to previous work with pure homeotropic anchoring \cite{topcolloids}, our genus-one colloids are engineered to have defects that are not only energetically favorable but also topologically required. We find that these hybrid washers tilt at an angle, with respect to the far-field direction $\vec{n}_0$, out of the horizontal plane in a homeotropically aligned cell, similar to the behavior of other colloids with hybrid anchoring \cite{tiltedhybrid}. Further, we introduce spherical colloids with homeotropic anchoring and show that they can be captured, creating a ``ring around the colloid'' configuration shown in Fig.\ref{Fig1}b. 
The presence of a spherical colloid inside a hybrid washer imposes axial symmetry causing the tilted hybrid washer to reorient parallel to the cell boundaries.

To study this system we disperse hybrid washers and homeotropic spherical colloids in 5CB-filled LC cells with homeotropic anchoring at the bounding surfaces. The liquid crystal used in all experiments is 4-cyano-biphenyl (5CB, Kingston Chemicals Ltd.). Washers are fabricated lithographically from SU-8 2005 negative tone photoresist (Microchem Corp.) using an OAI Model 100 mask aligner. Washers have an outer diameter of $20$ $\mu$m, an inner diameter of $10$ $\mu$m and thickness between $3-5$ $\mu$m. To achieve ``hybrid'' anchoring, we first sputter a $30$ nm film of chromium and treat all exposed surfaces with a 3\% wt.\ solution of dimethyloctadecyl[3-(trimethoxysilyl)propyl]ammonium chloride, 72\% (DMOAP, Sigma Aldrich) in a 9:1 mixture of ethanol:water prior to releasing washers from the substrate used for lithography \cite{GharbiSoftMatter}.This treatment imposes strong homeotropic anchoring on all faces except that in contact with the substrate, which has degenerate planar anchoring; hence, a hybrid washer. Homeotropic washers are fabricated from PDMS, which naturally imposes homeotropic anchoring without further treatment \cite{PDMS}. Washers are then released from the substrate and are suspended in 5CB along with similarly treated silica microspheres ($d_{s}$ = 5 $\mu$m, Bangs Laboratories). The suspension is introduced via capillarity into LC cells consisting of two DMOAP-treated glass slides separated by Mylar films with known thickness of $50$ $\mu$m. We observe our system using bright field and polarized optical microscopy with an upright microscope in transmission mode equipped with crossed polarizers (Zeiss AxioImager M1m) and a high resolution color camera (Zeiss AxioCam HRc). To manipulate particles we use optical tweezers driven by a $1065$ nm continuous wave fiber optic laser (IPG Photonics) with an output power of $2$ W and trap stiffness calibrated to be $0.04$ pN/nm in a water and sucrose solution.

When genus-one hybrid washers are placed in homeotropically aligned cells, they tilt with respect to the far-field direction $\vec{n}_0$ and induce the formation of a disclination ring around the washer's outer edge. This ring can be seen in the bright field and polarized optical microscopy images shown in Fig. 2a and 2b, respectively.
\begin{figure} [h]
\centering
\includegraphics[width=0.5\textwidth]{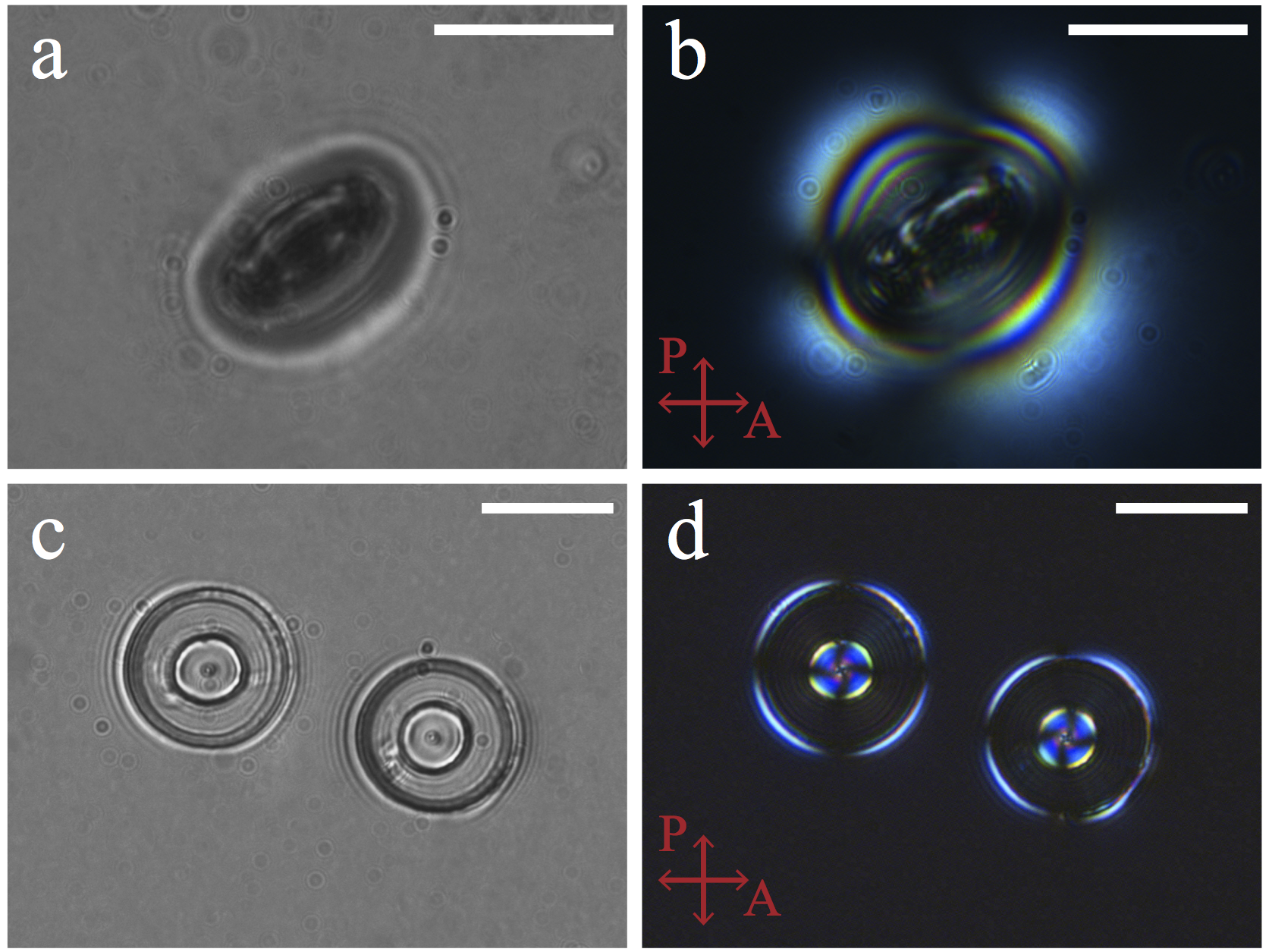}
\caption{(a) Bright field and (b) polarized optical microscopy images of hybrid washers in a homeotropic cell. We observe only an outer defect ring and a characteristic tilting of the charged washer. (c) Bright field and (d) polarized optical microscopy images of homeotropic washers in a homeotropically aligned cell. In this case, an inner point defect or defect loop forms to compensate for the outer defect loop. (All scale bars are $20$ $\mu$m.)} 
\label{Fig2} 
\end{figure} 
Whereas homeotropic genus-one washers lead to the formation of two defects, one inside and outside the washer (Fig.\ref{Fig2}c and \ref{Fig2}d)\cite{topcolloids}, the hybrid washers presented here are accompanied by only a single disclination. The difference stems from the fact that a loop encircling a handle of the washer measures a winding of the director by $2\pi$ for the homeotropic washer but only $\pi$ in the case of the hybrid washer. The hybrid washer is therefore topologically equivalent to the presence of a disclination ring whose profile has winding number $+1/2$ (with zero self-linking number). Such a ring carries a unit topological charge that must be balanced by a companion defect in the nematic, also carrying a charge of magnitude one, in the same way that a homeotropic spherical colloid is accompanied by either a hedgehog or a Saturn ring \cite{RMP2}. The disclination loop observed around the outside of the hybrid washer is therefore a topologically required defect. In contrast, a homeotropic washer is topologically equivalent to the presence of a disclination ring with winding number $+1$, but such a disclination can be replaced by a defect-free configuration through the ``escape in the third dimension" \cite{meyer1973existence}. Therefore, any defects accompanying the homeotropic washer must be present in charge-cancelling pairs, and are energetically beneficial but not topologically necessary  \cite{RMP2}. In this way, a change in surface anchoring on one face of the colloid alters the topology of the nematic director field that the colloid induces, including a change in the number of companion disclinations.
 
\begin{figure*}
\centering
\includegraphics[width=1\textwidth]{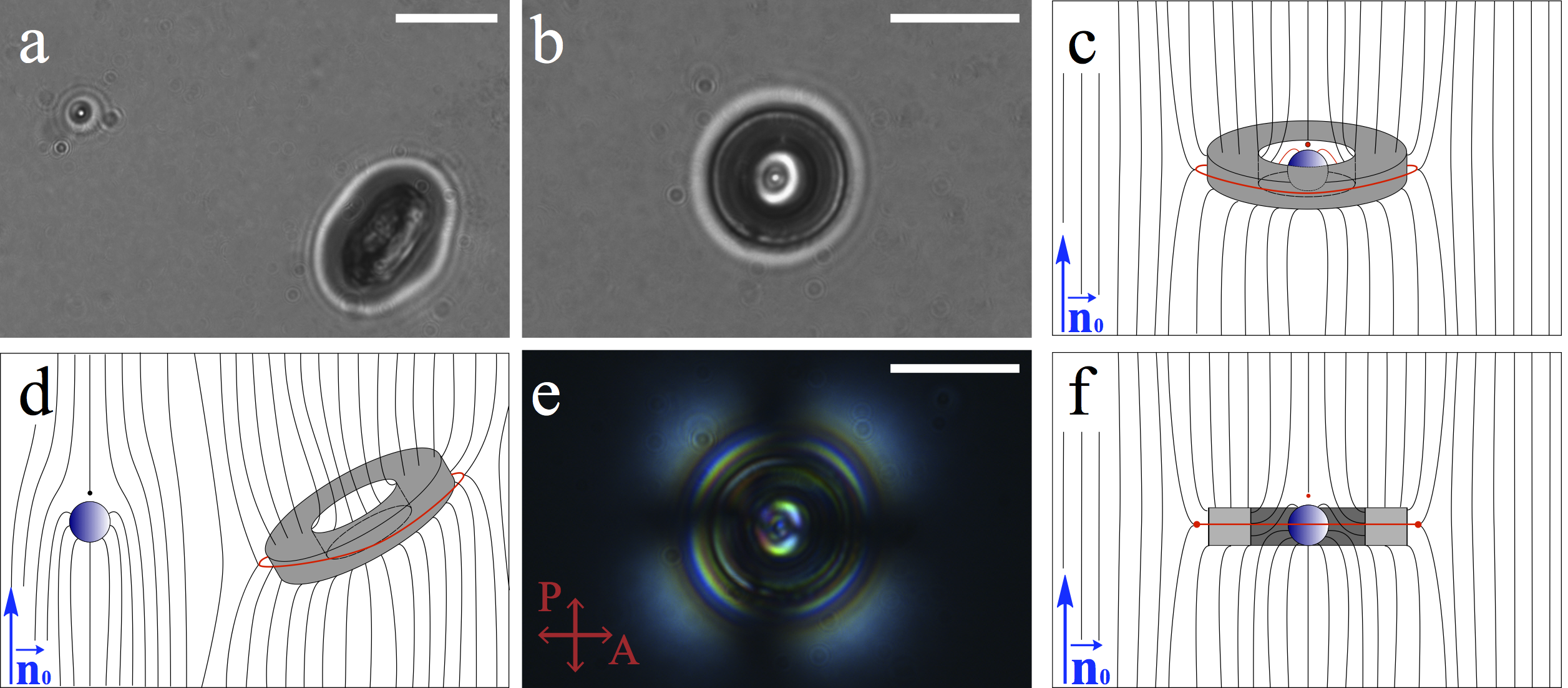}
\caption{A homeotropic spherical colloid (elastic dipole) and a nearby charged hybrid washer are shown in bright field in (a) and schematically in (d). Optical tweezers are then used to manipulate the washer to create the composite ring around the colloid assembly. The spherical colloid imposes axial symmetry allowing the composite to reorient perpendicular to the director field as seen in (b) bright field  and (e) polarized optical microscopy. (c,f) The spherical colloid is accompanied by a point defect and disfavors the hybrid washer from breaking symmetry, thereby enforcing a parallel orientation with respect to the far-field boundaries. (All scale bars are $20$ $\mu$m.)}
\label{Fig3}
\end{figure*}
The hybrid washer also exhibits remarkably different alignment in a homeotropic cell as compared with the homeotropic washer. The latter orients in the plane normal to $\vec{n}_0$. However, we observe the hybrid washer to consistently tilt out of this plane at a significant angle that depends on washer thickness, inner and outer diameters, and the thickness of the nematic cell. This tilt results from a competition between the elastic distortion near the degenerate planar face of the washer and that near the homeotropic face on its opposite side. If the washer were to lie in the horizontal plane, the director would bend by 90 degrees in going from the cell boundary to the washer's degenerate planar face, which would introduce a preference for tilting out of the horizontal plane and into a vertical plane. A washer with degenerate planar anchoring on both faces would be expected to rotate completely into a vertical plane. This preference to rotate out of plane competes with the homeotropic face's preference to lie in the plane normal to $\vec{n}_0$. The balance of these contributions to the elastic energy leads to a nonzero tilt of the colloid. 

On the other hand, we find that the hybrid washer can be induced to untilt by the addition of a spherical homeotropic colloid inside of it, as shown schematically in Fig.\ref{Fig1}b and Fig. \ref{Fig3}. The sphere is accompanied by its own companion defect, a hyperbolic hedgehog (Fig.\ref{Fig3}c and \ref{Fig3}f). Topological dipoles formed by spheres and hyperbolic hedgehogs prefer to orient parallel to the far-field director, with an axially symmetric director field nearby \cite{Lubensky1998}. Hence, this new topological pair spoils the tilted configuration preferred by the washer alone, instead forcing the washer to reorient in the horizontal plane to create an axially symmetric configuration. Since colloids are generally attracted to defect lines in bulk liquid crystal, we could not rely on spherical colloids to self-assemble inside hybrid washers because they prefer to aggregate along the outer ring induced by the washer. To overcome this barrier, we begin with a tilted hybrid washer and a spherical colloid in proximity to each other (see Fig.\ref{Fig3}a and \ref{Fig3}d) and we use optical tweezers, which serve two purposes: first, the heat due to the laser locally melts the LC close to the washer, thereby annihilating the outer defect loop; and second, once the LC is melted, we can position a spherical colloid inside the inner loop of the washer. When the optical trap is released, the sample cools down immediately and the spherical colloid becomes trapped inside the washer. The resulting untilted state of the washer is shown in experiment in Figs.\ref{Fig3}b and \ref{Fig3}e. In a similar fashion, we can separate the colloid from the washer to restore tilting. Note that the disclination around the outside of the washer remains present throughout this process, so the topology of the colloids is unchanged. Another possible configuration that is topologically equivalent would involve the hedgehog opening up into a Saturn ring around the spherical colloid, but we see no evidence of such a ring in the experimental data.

\begin{figure} [h]
\centering
\includegraphics[width=0.5\textwidth]{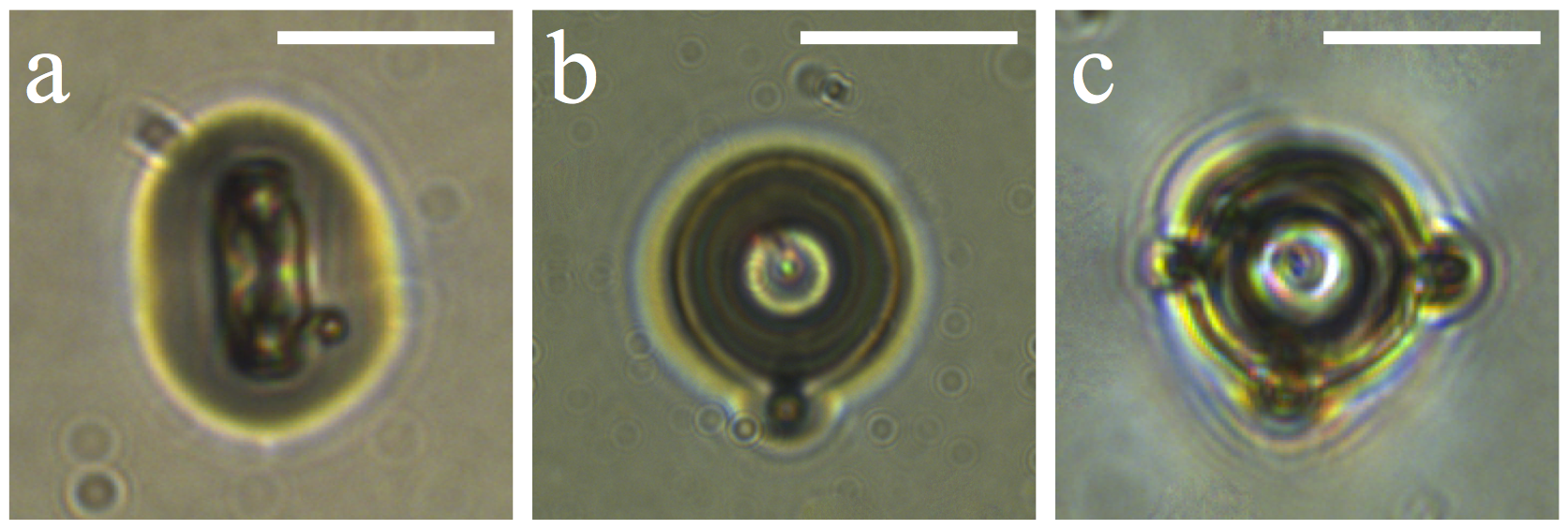}
\caption{(a) A single spherical colloid attracted to the washer's companion disclination loop does not alter the tilting. (b,c) Once a washer-colloid composite has been formed, it aligns perpendicular to the director field and subsequent colloids that collect along the outside loop do not re-establish the tilted configuration. (All scale bars are $20$ $\mu$m.)}
\label{Fig4} 
\end{figure}

The untilted configuration thus retains two defects in the nematic, one to balance the topological charge of the hybrid washer and one for the spherical colloid. The untilting is therefore not due to a change in topology, but rather a matter of elastic energy. In principle, it is topologically allowed for the sphere's companion defect and the washer's companion disclination ring to merge and escape into the third dimension, leaving the sphere and the washer as a topological pair. However, the two defects are spatially separated by the washer, so their approach is inhibited and energetically penalized. Although such a topological re-pairing was never observed, it may be possible by altering the geometries to bring the companion defects in greater proximity so they can pair and annihilate. It should also be noted that the aforementioned untilting behavior does not occur when spherical colloids are attracted solely to the outer disclination loop induced by a hybrid washer (Fig.\ref{Fig4}a). Finally, once a colloid has assembled \textit{inside} a hybrid washer, subsequent microspheres can assemble at the washer's disclination without re-establishing a tilted configuration (Figs. \ref{Fig4}b and \ref{Fig4}c). 

In conclusion, we have shown that with careful control of boundary conditions and particle shape we can engineer hybrid washers that are quite different from their purely homeotropic counterparts. Unlike homeotropic washers, hybrid washers are topologically charged and have a characteristic tilt when dispersed in a uniform director field. This observed tilting can be righted by the introduction of a homeotropic sphere inside the washer using the technique of optical tweezers. The trapping of the spherical particle in the washer's hole restores axial symmetry, creating a composite system that orients parallel to the far-field boundaries. We have shown how these components can be used to create complex particle aggregates by collecting individual beads at the ring around the washer, via long range elastic interactions. Thus, manipulation of colloid surface chemistry in addition to colloid shape greatly increases the range of organizing behaviors achievable by exploiting nematic elasticity and topology. This opens up exciting possibilities for an expanded library of hierarchical superstructures for applications such as metamaterials \cite{MusevicPRE2008,Stebe2009}.

The authors acknowledge support from the National Science Foundation (NSF) through the MRSEC Grant DMR11-20901 and thank the Isaac Newton Institute for their hospitality. This work was supported in part by a gift from The Mark Howard Shapiro and Anita Rae Shapiro Charitable Fund and by NSF Grant PHY11-25915 under the 2012 Kavli Institute for Theoretical Physics miniprogram ``Knotted Fields''. DAB was supported by the NSF through a Graduate Research Fellowship.

\footnotesize{
\bibliography{rsc} 
\bibliographystyle{rsc}}

\end{document}